\documentclass{amsart}
\usepackage{amssymb}
\usepackage{latexsym}
\date{\today}

\newcommand{\Z}{{\mathbb Z}}
\newcommand{\R}{{\mathbb R}}
\newcommand{\C}{{\mathbb C}}

\newtheorem{theorem}{Theorem}

\newtheorem{lemma}{Lemma}
\newtheorem{prop}{Proposition}

\renewcommand{\Im}{\mathrm{Im} \, }
\renewcommand{\Re}{\mathrm{Re} \, }
\newcommand{\tr}{\mathrm{tr} \, }

\def\e{\varepsilon}

\begin{document}
\title[Dynamical Lower Bounds on Wavepacket Spreading]{Scaling Estimates for Solutions and Dynamical Lower
Bounds on Wavepacket Spreading}

\author{David Damanik}

\address{Mathematics 253--37, California Institute of Technology, Pasadena, CA 91125,
USA}

\email{damanik@caltech.edu}

\author{Serguei Tcheremchantsev}

\address{UMR 6628--MAPMO, Universit\'{e} d'Orl\'{e}ans, B.P.~6759, F-45067 Orl\'{e}ans Cedex,
France}

\email{serguei.tcherem@labomath.univ-orleans.fr}

\thanks{D.\ D.\ was supported in part by NSF grant DMS--0227289.}

\begin{abstract}
We establish quantum dynamical lower bounds for discrete one-dimensional Schr\"odinger
operators in situations where, in addition to power-law upper bounds on solutions
corresponding to energies in the spectrum, one also has lower bounds following a scaling
law. As a consequence, we obtain improved dynamical results for the Fibonacci Hamiltonian
and related models.
\end{abstract}

\maketitle

\section{Introduction}

Consider a discrete one-dimensional Schr\"odinger operator,
\begin{equation}\label{oper}
[H \psi] (n)=\psi(n+1)+\psi(n-1) + V(n)\psi(n),
\end{equation}
on $\ell^2(\Z)$ or $\ell^2(\Z_+)$, where $\Z_+ = \{1,2,\ldots\}$. In the case of
$\ell^2(\Z_+)$, we will work with a Dirichlet boundary condition, $\psi(0) = 0$, but our
results easily extend to all other self-adjoint boundary conditions.

A number of recent papers (e.g., \cite{DST,DT,JL,JL2,JSS,KKL,T2,T}) were devoted to
proving lower bounds on the spreading of an initially localized wavepacket, say $\psi =
\delta_1$, under the dynamics governed by $H$, typically in situations where the spectral
measure of $\delta_1$ with respect to $H$ is purely singular and sometimes even pure
point.

A standard quantity that is considered to measure the spreading of the wavefunction is
the following: For $p > 0$, define
\begin{equation}\label{mpo}
\langle |X|_{\delta_1}^p \rangle (T) = \sum_n |n|^p a(n,T),
\end{equation}
where
\begin{equation}\label{ant}
a(n,T)=\frac{2}{T} \int_0^{+\infty} e^{-2t/T} | \langle e^{-itH} \delta_1, \delta_n
\rangle |^2 \, dt.
\end{equation}
Clearly, the faster $\langle |X|_{\delta_1}^p \rangle (T)$ grows, the faster $e^{-itH}
\delta_1$ spreads out, at least averaged in time. One typically wants to prove power-law
lower bounds on $\langle |X|_{\delta_1}^p \rangle (T)$ and hence it is natural to define
the following quantity: For $p > 0$, define the lower growth exponent
$\beta^-_{\delta_1}(p)$ by
$$
\beta^-_{\delta_1}(p)=\liminf_{T \to +\infty} \frac{{\rm log} \, \langle |X|_{\delta_1}^p
\rangle (T) }{ {\rm log} \, T}.
$$

When one wants to bound these exponents from below for specific models, it is useful to
connect these quantities to the qualitative behavior of the solutions of the difference
equation
\begin{equation}\label{eve}
u(n+1) + u(n-1) + V(n)u(n) = E u(n)
\end{equation}
for energies $E$ in the spectrum of the operator $H$. In fact, most of the known results
are based on such a correspondence; compare \cite{JL,JL2} for an approach where the link
is furnished by Hausdorff-dimensional properties of spectral measures, and \cite{DST,DT}
for a direct approach without intermediate step. The latter papers use power-law upper
bounds on solutions corresponding to energies from a set $S$ to derive lower bounds for
$\beta^-_{\delta_1}(p)$. The set $S$ can even be very small. One already gets non-trivial
bounds when $S$ is not empty. If $S$ is not negligible with respect to the spectral
measure of $\delta_1$, the bounds are stronger, but there are situations of interest
(e.g., random polymer models \cite{JSS}), where the spectral measure assigns zero weight
to $S$.

While both approaches yield bounds on $\beta^-_{\delta_1}(p)$ for all $p>0$, in concrete
applications there is a transition point, $p_0$, such that the method from \cite{JL,JL2}
works better for $0 < p < p_0$, whereas the method from \cite{DST,DT} gives better bounds
for $p > p_0$.

Our goal here is to develop an approach that, whenever it applies, gives stronger lower
bounds than both previous methods throughout the entire range of the powers $p$.

A model for which the exponents $\beta^-_{\delta_1}(p)$ have been heavily studied (e.g.,
\cite{d,dkl,DST,DT,JL2,KKL}) is given by the Fibonacci Hamiltonian. This is the standard
model of a one-dimensional quasicrystal and it is one of the few for which one can
actually prove ``anomalous'' transport properties rigorously; compare the discussion in
\cite{KKL}. With this model in mind, we will refine the results from \cite{DST,DT,KKL} in
what follows. It turns out that the bounds on $\beta^-_{\delta_1}(p)$ can be considerably
strengthened if, in addition to power-law upper bounds for solutions, one also assumes
suitable lower bounds. The necessary input does in fact hold for the Fibonacci model and
related ones, as we will show, and we thereby obtain improved dynamical results that are
strictly better than all previously known ones. While we will discuss this issue in more
detail later, we mention at this point that, for this particular model, the paper
\cite{KKL} had the best previous bounds for small values of $p$, while for large values
of $p$, the best previous bounds were obtained in \cite{DST,DT}. Here we will get
stronger bounds for all values of $p$.

Let us now specify the assumptions we are going to work with. We consider real solutions
$u$ of the difference equation \eqref{eve}. If
\begin{equation}\label{norm}
|u(0)|^2+|u(1)|^2=1,
\end{equation}
we say that $u$ is normalized. For $L \ge 1$, we define
$$
\|u\|_L^2 = \sum_{n=1}^{[L]} |u(n)|^2 + (L-[L]) |u([L]+1)|^2.
$$
We assume that for some non-empty $A \subseteq \R$, the following conditions are
satisfied:
\begin{itemize}
\item[(a)] There exist constants $C, \alpha>0$ such that for every $E \in A$, every
normalized solution $u$ of \eqref{eve}, and every $L \ge 1$,
$$
||u||_L^2 \le C L^{2\alpha+1}.
$$
\item[(b)] There exist constants $0 < k,\gamma <1$ and $L_0 \ge 1$ such that for every $E
\in A$, every solution $u$ of \eqref{eve}, and every $L \ge L_0$,
$$
||u||_L^2 \ge (1 + \gamma) ||u||_{kL}^2.
$$
\end{itemize}
It is easy to see that (b) implies the following:
\begin{itemize}
\item[(c)] There exist constants $D, \kappa > 0$ such that for every $E \in A$, every
normalized solution $u$ of \eqref{eve}, and every $L \ge 1$,
$$
||u||_L \ge D L^\kappa.
$$
\end{itemize}
We make this condition explicit since the constant $\kappa$ is crucial in the dynamical
lower bounds we will prove below, that is, it is desirable to find the largest possible
value of $\kappa$ such that (c) holds. In our general treatment we will only assume the
conditions (a) and (b), but in concrete applications one can try to optimize $\kappa$.

Denote by $F$ the Borel transform of the spectral measure $\mu$ associated with the
operator $H$ and vector $\delta_1$. That is,
\begin{equation}\label{borel}
F(z)=\langle (H-z)^{-1} \delta_1, \delta_1 \rangle = \int_{\R} \frac{d \mu (x)}{x-z}.
\end{equation}
For $\Delta \subset \R$ and $\nu > 0$, we let
$$
I(\Delta, \nu)=\nu \int_\Delta (\Im F(E+i\nu))^2 \, dE
$$
and write $\Delta_\nu$ for the $\nu$-neighborhood of $\Delta$.

Our first result establishes a lower bound for $\langle |X|_{\delta_1}^p \rangle (T)$ in
terms of the integrals $I(A_{2 \e}, \e)$, where $\e = 1/T$. We stress that (a)--(c) above
only concern the behavior of the solutions on the right half-line, even in the case of a
whole-line operator. This is natural from a physical point of view, for if there is
``transport'' on a half-line, there should also be ``transport'' for the whole-line
model. On a mathematical level, this intuition does not always translate into an easy
proof and some earlier papers needed certain symmetry assumptions to prove quantum
dynamical lower bounds for whole-line operators (e.g., \cite{d,JL2}). However, the
Jitomirskaya-Last theory \cite{JL,JL2} does allow for a decent whole-line version that
works with solution estimates on one half-line only; see \cite{dkl}.

\begin{theorem}\label{hp}
Let $H$ be a discrete Schr\"odinger operator, given by \eqref{oper}, acting on
$\ell^2(\Z)$ or $\ell^2(\Z_+)$. Assume that the conditions {\rm (a)} and {\rm (b)} are
satisfied for some set $A \subseteq \R$. Then, for $0 < p \le 2 \alpha +1$,
\begin{equation}\label{fa1}
\langle |X|_{\delta_1}^p \rangle (T) \gtrsim T^{2(p+2 \kappa)/(2 \alpha +1+2\kappa)}
I(A_{2 \e}, \e),
\end{equation} whereas for $p> 2\alpha+1$, we have
\begin{equation}\label{fa2}
\langle |X|_{\delta_1}^p \rangle (T) \gtrsim T^2 I(A_{2 \e}, \e)
\end{equation}
and
\begin{equation}\label{fa3}
\langle |X|_{\delta_1}^p \rangle (T) \gtrsim T^{\frac{p+2\kappa}{2\alpha+1}} I(A_{2\e},
\e).
\end{equation}
The bounds \eqref{fa1}--\eqref{fa3} hold for $T \ge T_0$ with $\e=1/T$.
\end{theorem}

\noindent\textit{Remark.} We write $f(T) \gtrsim g(T)$ if there is a positive,
$T$-independent constant $C$ such that $f(T) \ge C g(T)$. In the bounds above, these
constants depend on the values of $p,C,\alpha,D,\kappa,k,\gamma$ in (a)--(c). They are
given by
$$
\frac{\gamma k^p}{12 \pi} \, \left( \frac{\gamma^2}{5184 \, C} \right)^\frac{p+2
\kappa}{2 \alpha + 1 + 2 \kappa} D^{2- \frac{2(p+2\kappa)}{2\alpha+1+2 \kappa}}
$$
in \eqref{fa1}--\eqref{fa2} and
$$
\frac{\gamma k^p}{12\pi} \left( \frac{\gamma^2}{5184 \, C^2} \right)^\frac{p + 2 \kappa}{2}
D^2
$$
in \eqref{fa3}, respectively.

\medskip

In order to apply this theorem, we need to establish a lower bound for $I(A_{2\e}, \e)$.
In general, this is a difficult problem. One can show that if $\mu (A)=1$, then
$$
\e \int_{A_{2\e}} (\Im F(E + i\e))^2 \, dE \sim \e \int_{\R} (\Im F(E+i\e))^2 \, dE,
$$
where the last integral over $\R$ is closely related to the time-averaged return
probability and the correlation dimension of the spectral measure \cite{T2,T}.

If $\mu (A_{\e})>0$,
$$
I(A_{2\e}, \e) \gtrsim \e \int_{A_{2\e}} \Im F(E + i\e) \, dE \gtrsim \e \mu (A_{\e})
$$
(for a proof of the last inequality, see, e.g., \cite{KKL}). In particular, if
$\mu(A)>0$, then $I (A_{2\e}, \e) \gtrsim \e $. However, such a lower bound for $I$ is
not optimal in most cases of interest.

When considering the time-averaged moments, a method combining the Parseval formula and
the classical Guarneri approach was proposed in \cite{T}. It yields better lower bounds
without any information on $I$. As a particular consequence of this method, we can derive
the following result:

\begin{theorem}\label{combi}
Let $H$ be a discrete Schr\"odinger operator, given by \eqref{oper}, acting on
$\ell^2(\Z)$ or $\ell^2(\Z_+)$. Suppose $A$ is a set {\rm (}possibly depending on $T${\rm
)} such that $\mu (A)>0 $ and
\begin{equation}\label{fa30}
\langle |X|_{\delta_1}^p \rangle (T) \ge g_p(T) I(A_{2\e}, \e), \; \e=\frac1T, \;p>0,
\end{equation}
where $g_p(T)$ are positive functions of $T$. Then
$$
\langle |X|_{\delta_1}^p \rangle (T) \gtrsim (g_p(T))^{p/(p+1)} (\mu (A))^{(1 +
2p)/(p+1)}.
$$
In particular, if the set $A$ does not depend on time, then
$$
\langle |X|_{\delta_1}^p \rangle (T) \gtrsim (g_p(T))^{p/(p+1)}.
$$
\end{theorem}

This theorem, combined with Theorem~\ref{hp}, allows us to derive the following lower
bound for $\beta^-_{\delta_1}(p)$:

\begin{theorem}\label{mome}
Let $H$ be a discrete Schr\"odinger operator, given by \eqref{oper}, acting on
$\ell^2(\Z)$ or $\ell^2(\Z_+)$. Assume $A \subseteq \R$ is such that $\mu (A)>0$ and
conditions {\rm (a)} and {\rm (b)} hold. Then,
\begin{equation}\label{ourbound}
\beta^-_{\delta_1}(p) \ge \begin{cases} \frac{p (p+2\kappa)}{(p+1)(\alpha + \kappa + 1/2)} & p \le 2 \alpha+1, \\
\frac{p}{\alpha+1} & p>2 \alpha +1. \end{cases}
\end{equation}
\end{theorem}

\begin{proof}
The inequality for $0 < p \le 2 \alpha + 1$ follows directly from Theorems~\ref{hp} and
\ref{combi}. The Jensen inequality implies that $(\langle |X|^p_{\delta_1} \rangle
(T))^{1/p}$ is non-decreasing in $p$. Taking $p_0=2 \alpha+1$ and $p>p_0$, we get the
inequality for $p > 2 \alpha + 1$.
\end{proof}

\noindent\textit{Remarks.} (i) Since $2 \kappa \le 2\alpha+1$, an application of
Theorem~\ref{combi} to \eqref{fa2} or \eqref{fa3} does not give better bounds for
$p>2\alpha+1$.\\[1mm]
(ii) If $\mu(A) > 0$, it follows from the existence of generalized eigenfunctions
\cite{b,Si1} that the constant $\kappa$ in (c) cannot be larger than $1/2$.
\\[1mm]
(iii) The Jitomirskaya-Last approach \cite{JL,JL2} (see also \cite{dkl,KKL}) gives the
following effective way to obtain dynamical bounds from solution estimates: Assuming
conditions (a) and (c) for a set $A$ with $\mu(A) > 0$, one obtains that
\begin{equation}\label{jltbound}
\beta^-_{\delta_1}(p) \ge \frac{2p\kappa}{\alpha + \kappa + 1/2}.
\end{equation}
If we assume (b) rather than (c), we can prove the bound \eqref{ourbound}. It is easy to
check that \eqref{ourbound} coincides with \eqref{jltbound} when $\kappa = 1/2$ and
\eqref{ourbound} is strictly stronger than \eqref{jltbound} for every $p > 0$ when
$\kappa < 1/2$. Thus, by Remark~(ii) above, our bounds are favorable to the ones obtained
through the Jitomirskaya-Last approach in all situations where one has (a) and (b). While
in general (b) is a stronger condition than (c), in concrete applications one often
really proves (b) in order to show
(c) (e.g., in our and previous studies of the Fibonacci Hamiltonian and related models).\\[1mm]
(iv) The paper \cite{DST} worked under the sole assumption~(a) and derived the bound
$$
\beta^-_{\delta_1}(p) \ge \frac{p-3 \alpha}{\alpha + 1}
$$
for all $p > 0$. In particular, this statement is vacuous when $p \le 3 \alpha$. Thus,
under the additional assumption~(b), Theorem~\ref{mome} extends the range of relevant $p$
to the entire interval $(0,\infty)$ and on top of that improves the lower bound on
$\beta_{\delta_1}^-(p)$. Moreover, even if we only assume condition (a) for some set $A$
with $\mu(A) > 0$, we can improve the bounds from \cite{DST} by applying Theorem~\ref{combi}.\\[1mm]
(v) It is possible to derive lower bounds for the outside probabilities,
$$
P (|n| \ge K(T), T) = \sum_{|n| \ge K(T)} a(n,T),
$$
from this result. Here, $K(T)$ is a suitable growing function of $T$. This is discussed
in Section~\ref{outprob}.

\bigskip

We will demonstrate how to apply our general dynamical results to operators with Sturmian
potentials on $\Z$, that is,
\begin{equation}\label{sturmpot}
V(n) = \lambda \chi_{[1-\omega,1)}(n \omega \mod 1), \; n \in \Z,
\end{equation}
with coupling constant $\lambda > 0$ and irrational rotation number $\omega \in (0,1)$.
If $\omega = (\sqrt{5}-1)/2$, then $V$ is usually called the Fibonacci potential and the
associated operator is called the Fibonacci Hamiltonian.

Consider the continued fraction expansion of $\omega$,
$$
\omega = \cfrac{1}{a_1+ \cfrac{1}{a_2+ \cfrac{1}{a_3 + \cdots}}}
$$
with uniquely determined $a_n \in \Z_+$. The number $\omega$ is said to have bounded
partial quotients if the sequence $\{a_n\}$ is bounded. In this case,
\begin{equation}\label{domega}
d(\omega) = \limsup_{n \rightarrow \infty} \frac{1}{n} \sum_{k=1}^n a_k < \infty.
\end{equation}
The set of such numbers $\omega$ is uncountable but has Lebesgue measure zero. Note that
$\omega = (\sqrt{5}-1)/2$ is contained in this set since, in this case, $a_n = 1$ for
every $n$.

\begin{theorem}\label{fibthm}
Suppose $\lambda > 0$ and $\omega \in (0,1)$ is irrational with $a_n \le C$. Consider the
operator \eqref{oper} with potential given by \eqref{sturmpot}. With
$$
\alpha = D \, d(\omega) \, \log (2 + \sqrt{8 + \lambda^2})
$$
{\rm (}$D$ is some universal constant{\rm )} and
$$
\kappa = \frac{\log (\sqrt{17} / 4)}{(C+1)^5},
$$
the following dynamical bounds hold true:
\begin{equation}\label{fibbetabound}
\beta^-_{\delta_1}(p) \ge \begin{cases} \frac{p (p+2\kappa)}{(p+1)(\alpha + \kappa + 1/2)} & p \le 2 \alpha+1, \\
\frac{p}{\alpha+1} & p>2 \alpha +1. \end{cases}
\end{equation}
\end{theorem}

Let us compare these bounds with the ones that were previously known (the papers
\cite{d,dkl,dl,DST,DT,JL2,KKL} prove dynamical results for Sturmian potentials).

We first consider the Fibonacci case, that is, $\omega = (\sqrt{5}-1)/2$.\footnote{As
discussed in Section~\ref{fibsec}, the value of $\kappa$ can often be slightly improved
over what is stated above. In particular, Killip et al.\ can work with $\kappa = \log [
\sqrt{17} / (20 \log \omega^{-1}) ]$ in the Fibonacci case. For this particular case, we
can use this value also in \eqref{fibbetabound}.} For small values of $p > 0$, the best
previously known bound was obtained in \cite{KKL}, using the Jitmorskaya-Last approach,
and consequently reads
$$
\beta^-_{\delta_1}(p) \ge \frac{2p\kappa}{\alpha + \kappa + 1/2}.
$$
This bound is valid for all values of $p > 0$, but for $p$ large, the bound
\begin{equation}\label{dtbound}
\beta^-_{\delta_1}(p) \ge \frac{p-3 \alpha}{\alpha + 1},
\end{equation}
obtained in \cite{DST}, which also holds for all $p > 0$, is better. As discussed in
Remarks~(iii) and (iv) after Theorem~\ref{mome}, the bound \eqref{fibbetabound} is
strictly better than both of these bounds for all $p > 0$.

For other $\omega$'s with bounded partial quotients, the gap between our bound and
previously known bounds is even bigger. For small $p>0$, the best bound was \cite{d} (see
also \cite{dkl})
\begin{equation}\label{sturmdkl}
\beta^-_{\delta_1}(p) \ge \frac{2 p \kappa_\lambda}{\alpha + \kappa_\lambda + 1/2}
\end{equation}
with some small $\lambda$-dependent $\kappa_\lambda > 0$ which obeys $\kappa_\lambda \to
0$ as $\lambda \to \infty$, whereas for larger values of $p$, it is again better to use
the bound \eqref{dtbound}. Again we see that \eqref{fibbetabound} improves upon all
previously known dynamical bounds for these potentials.

To summarize, we establish a new effective way to derive quantum dynamical lower bounds
from solution estimates. As with previous methods, we require suitable upper and lower
bounds on solutions to the difference equation associated with the operator. While the
upper bounds we need are the same as in previous approaches, the lower bounds are
slightly stronger. Whenever our result applies, it gives better dynamical results (except
in extreme cases, where the derived bounds are the same). The particular case of the
Fibonacci Hamiltonian, and the related Sturmian models, is discussed in detail.

The question of how to obtain any upper bounds on the fast part of the time evolution
remains an important open problem. For example, it is in general not clear how to bound
$\langle |X|_{\delta_1}^p \rangle (T)$ or $\beta^-_{\delta_1}(p)$ from above. (The only
exception is the case of growing sparse potentials \cite{cm,T}.)

The organization of the paper is as follows. We prove Theorem~\ref{hp} for half-line and
whole-line operators in Sections~\ref{ndlb} and \ref{ndlbw}, respectively. Then we prove
Theorem~\ref{combi} in Section~\ref{combiproof} and discuss some consequences of
Theorem~\ref{mome} for outside probabilities in Section~\ref{outprob}. Finally, we prove
Theorem~\ref{fibthm} in Section~\ref{fibsec} and present some simplifications and
improvements of some central results within the Jitomirskaya-Last theory in the appendix.

\section{Proof of Theorem~\ref{hp} For Half-Line Operators}\label{ndlb}

In this section we consider Schr\"odinger operators on the half-line. We first introduce
notation and gather a few useful results. Then we prove Theorem~\ref{hp} for half-line
operators.

Let $H$ be a discrete Schr\"odinger operator on $\ell^2(\Z_+)$ with potential $V$;
compare \eqref{oper}. For $z \in \C$ and $\theta \in [0, 2 \pi)$, denote by
$u_\theta(n,z)$ the solution to the equation
\begin{equation}\label{evez}
u(n+1) + u(n-1) + V(n)u(n) = z u(n)
\end{equation}
that obeys $u_\theta (0,z) = \sin \theta$, $u_\theta(1,z) = \cos \theta$. With $F$ from
\eqref{borel}, one can verify \cite{JL} that for any $n \ge 1$,
\begin{equation}\label{eq1}
u(n,z) := \langle (H-z)^{-1} \delta_1, \delta_n \rangle = F(z) u_0(n,z) -u_{\pi/2}(n,z).
\end{equation}

Thus,
\begin{equation}\label{eq1a}
(u(n+1,z), u(n,z))^T=T(n,0; z) (F(z), -1)^T, \ n \ge 1,
\end{equation}
where $T$ is the transfer matrix associated to the equation $Hu=zu$:
$$
T(n,0;z)=\left( \begin{array}{cr} u_0(n+1,z) & u_{\pi/2}(n+1,z) \\
u_0(n,z) & u_{\pi/2} (n,z)\end{array} \right).
$$

If one has good control of the functions $u(n,z)$ for complex $z$, then two kinds of
results can be obtained.

The first is related with the study of the function $F(z)$. In particular, since $\mu
([E-\e, E+\e]) \le 2 \e \, \Im F(E+i\e)$, an upper bound for the measure of intervals
follows from an upper bound on $\Im F(z)$. Such a bound provides a lower bound for the
lower Hausdorff or packing dimension of the spectral measure. On the other hand, lower
bounds on $|F(z)|$ can be used to show singularity of the spectral measure; compare
\cite{JL}.

The second group of results is based on the Parseval formula. For any $f$, define
$$
\langle f(t) \rangle (T)= \frac{2}{T} \int_0^{\infty} e^{-2t/T} f(t) dt.
$$
Then (cf.~\cite{KKL,rs4})
$$
\langle | \langle e^{-itH} \psi, \delta_n\rangle |^2 \rangle (T)= \frac{\e}{\pi} \int_\R
|\langle (H - E - i\e)^{-1} \psi, \delta_n \rangle |^2 dE,
$$
where $\psi \in \ell^2(\Z_+)$ and $\e=1/T$. In particular, for $a(n,T)$ as defined in \eqref{ant},
we obtain
\begin{equation}\label{pars}
a(n,T) = \frac{\e}{\pi} \int_\R |u(n,E+i\e)|^2 dE.
\end{equation}
This formula can be used to bound $a(n,T)$ from below or from above.

To apply the formula \eqref{pars} directly, one should have good control of
$|u(n,E+i\e)|$ depending on $n,E, \e$ for small values of $\e$. However, in most
applications, it is much easier to obtain information on solutions to the equation
\eqref{evez} with real $z=E$. (Sparse potentials represent a rare case where solutions
with complex $z$ can be studied directly; see \cite{T}.) Therefore, one uses perturbative
methods to link solutions to the equation \eqref{evez} with real $z=E$ and complex
$z=E+i\e$ (with small~$\e$). This also allows one to study the Borel transform $F(z)$ of
the spectral measure \cite{JL}.
\par
Probably the most obvious approach is based on upper bounds on the norm $||T(n,0;z)||$.
If one has such a bound for $||T(m,0; E)||$, $m \le n$, one can apply the method of Simon
\cite{Si} to get bounds for complex $z$ (if $n$ is not too large). Since $\det T=1$,
\eqref{eq1a} implies that for $n \ge 1$,
\begin{equation}\label{eq3}
|u(n+1,z)|^2 +|u(n,z)|^2\ge ||T(n,0; z)||^{-2} (|F(z)|^2+1).
\end{equation}

Therefore, if one has a non-trivial upper bound for $||T(n,0; E)||$ on some set of
energies $E$, \eqref{pars} and \eqref{eq3} yield a lower bound for the quantity
$a(n+1,T)+a(n,T)$, and thus a lower bound for the time-averaged moments
$$
\langle |X|^p \rangle (T)=\sum_n n^p a(n,T), \ p>0,
$$
and outside probabilities
$$
\sum_{n \ge M} a(n,T)
$$
(with $M$ depending on $T$) \cite{DST,DT}. Although this method gives good lower bounds
for moments with large $p$, it is clearly not optimal since the left-hand side of
\eqref{eq3} may be much larger than what the perturbative argument gives as a lower bound
for the right-hand side (at least, for some values of $E$).


The upper bound on $||T(n,0; z)||$ can be also used to bound $\mathrm{Im} \, F(z)$ from
above. Since
\begin{align*}
\mathrm{Im} \, F(z) & = \e ||R(z)\delta_1||^2 \\
& = \e \sum_n |u(n,z)|^2 \\
& \ge \frac{\e}{2} (|F(z)|^2+1) \sum_n ||T(n,0;z)||^{-2} \\
& \ge \frac{\e}{2} \left(\mathrm{Im} \, F(z)\right)^2 \sum_n ||T(n,0; z)||^{-2},
\end{align*}
it follows that
$$
\Im F(z) \le \frac{2}{\e} \left( \sum_n ||T(n,0;z)||^{-2} \right)^{-1}.
$$

However, in most applications, better bounds can be obtained using the Jitomirskaya-Last
approach discussed next. This method was proposed in \cite{JL} and later developed in
\cite{KKL}. We will discuss certain improvements of this theory in the appendix. Let
$E',E \in \R$, $\e>0$, and $z=E'+i \e$. The starting point is the following formula
\cite{JL,KKL}:
\begin{equation}\label{eq2}
u(n,z)=F(z)u_0(n,E)-u_{\pi/2} (n,E)+(z-E) (K(E)u(z))(n), \ n \ge 1,
\end{equation}
where
$$
(K(E) \psi) (n)=\sum_{j=1}^n K(n,j,E)\psi (j)
$$
and
$$
K(n,j,E) = u_0(n,E)u_{\pi/2}(j,E) - u_{\pi/2}(n,E) u_0(j,E).
$$
For $L \ge 1$, consider the ($[L]$ or $[L] + 1)$)-dimensional space with the inner
product
$$
\langle f,g \rangle =\sum_{n=1}^{[L]} f(n){\overline {g(n)}} +(L-[L])
f([L]+1){\overline {g([L]+1)}}.
$$
We will denote the corresponding norm by $||f||_L$. The Hilbert-Schmidt norm of $K(E)$ in
this space is given by \cite{JL, KKL}:
\begin{align}\label{eq30}
|||K(E)|||_L^2 &= ||u_0(E)||_L^2 \, ||u_{\pi/2}(E)||_L^2 -
 \langle u_0(E), u_{\pi/2}(E) \rangle _L ^2 \\
& = \sup_\theta ||u_\theta (E)||_L^2 \inf_\theta ||u_\theta (E)||^2 \nonumber.
\end{align}
Let
$$
u(n,E) := F(z) u_0(n,E) - u_{\pi/2} (n,E).
$$
Assume that $|z-E| \le \delta$. Since the operator norm does not exceed the
Hilbert-Schmidt norm, it follows from \eqref{eq2} that for any $L$,
\begin{equation}\label{eqa3}
(1+\delta |||K(E)|||_L)^{-1}||u(E)||_L \le ||u(z)||_L \le ( 1-\delta |||K(E)|||_L )^{-1}
||u(E)||_L
\end{equation}
(the second inequality holds under the
condition $\delta |||K(E)|||_L<1$). Thus, if the norm $|||K(E)|||_L$ is not too large, in
order to control $||u(z)||_L$, it is sufficient to control $||u(E)||_L$. Since both
solutions $u_0(n,E), \ u_{\pi/2} (n,E)$ are real,
\begin{equation}\label{eqa4}
|u(n,E)|^2=(\Re F(z) u_0(n,E) - u_{\pi/2} (n,E))^2 + (\Im F(z) u_0(n,E))^2.
\end{equation}
Having some information about solutions $u_0(n,E), \ u_{\pi/2} (n,E)$, we have some
control of $||u(z)||_L$. In particular, one can prove bounds on $|F(z)|$ or $\mathrm{Im}
\, F(z)$. On the other hand, a lower bound on $||u(z)||_L$ for $E$ from some set yields,
via \eqref{pars}, a lower bound for the inside probabilities
$$
\sum_{n \le L} a(n,T).
$$
Results of this kind were obtained in \cite{JL,KKL}. As noted above, we present
simplified and unified proofs of some of them in the appendix.

\medskip

We are now ready to give the

\begin{proof}[Proof of Theorem~\ref{hp} for half-line operators.]
We shall estimate from below the quantities $$h_p(E'+i\e)=\sum_n n^p |u(n,E'+i\e)|^2$$
with $E' \in A_{2\e}$. It follows from (a) and \eqref{eq30} that, for every $E \in A$,
\begin{equation}\label{fo4}
||K(E)||_L^2 \le |||K(E)|||_L^2 \le C L^{2 \alpha +1} f(L,E),
\end{equation}
where $|| \cdot ||_L$ is the operator norm and
$$
f(L,E)=\inf_\theta ||u_\theta (n,E)||_L^2.
$$
Let $E' \in A_{2\e}$. That is, there exists $E \in A$ such that $|E'-E|\le 2\e$. Define
$z=E'+i\e$. The bound \eqref{eqa3} holds with $\delta=3 \e$. From now on we shall assume
that
\begin{equation}\label{betab}
\beta := 3 \e |||K(E)|||_L \le 1/4.
\end{equation}
By \eqref{fo4}, this holds provided that $9 \e^2 C L^{2 \alpha+1 } f(L,E) \le 1/16$. The
latter condition is satisfied if $L$ is not too large (depending on $\e,E$).

Let $0<M<L$. We can estimate from below, using \eqref{eqa3},
\begin{align*}
P(M,L) & := ||u(z)||_L^2-||u(z)||_M^2 \\
& \ge (1+\delta ||K(E)||_L)^{-2} ||u(E)||_L^2 - (1-\delta ||K(E)||_M)^{-2}||u(E)||_M^2 \\
& \ge (1+\delta ||K(E)||_L)^{-2} ||u(E)||_L^2 - (1-\delta ||K(E)||_L)^{-2}||u(E)||_M^2
\end{align*}
since $||K(E)||_M \le ||K(E)||_L$. Thus, by \eqref{betab},
\begin{align}\label{fo7}
P(M,L) & \ge \frac{1}{(1+\beta)^2} \sum_{n=M+1}^L |u(n,E)|^2 + \left(
(1+\beta)^{-2}-(1-\beta)^{-2} \right)||u(E)||_M^2 \\
& \ge \frac{1}{2} \sum_{n=M+1}^L |u(n,E)|^2 - 8 \beta ||u(E)||_M^2. \nonumber
\end{align}
The identity \eqref{eqa4} implies
$$
|u(n,E)|^2 = (w (n,E))^2 + (\Im F(z) u_0(n,E))^2,
$$
where $w(n)=\Re F(z) \, u_0(n,E) - u_{\pi/2}(n,E)$ is a real solution to the equation
\eqref{eve}. Thus, \eqref{fo7} implies
\begin{align}
\label{fo8} P(M,L) \ge \frac{1}{2} & \sum_{n=M+1}^L (w (n,E))^2 - 8 \beta ||w(E)||_M^2 \\
& + (\Im F(z))^2 \left( \frac{1}{2} \sum_{n=M+1}^L (u_0(n,E))^2 - 8 \beta ||u_0(E)||_M^2
\right). \nonumber
\end{align}

Let $k$ be the constant from condition (b). If $\beta \le \gamma /24$, we get for every
real solution $v$ of \eqref{eve},\footnote{Here, we need $L$ large enough, that is,
$\varepsilon$ small enough or, in other words, $T$ large enough.}
\begin{align}\label{kboundg}
\frac{1}{2} \sum_{n = kL + 1}^L  (v(n,E))^2 - 8 \beta \sum_{n=1}^{kL} (v(n,E))^2 & \ge
\left( \frac{1}{2} - \frac{8\beta}{\gamma} \right)
\sum_{n = kL + 1}^L (v (n,E))^2 \\
& \ge \frac{1}{6} \sum_{n = kL + 1}^L (v(n,E))^2 \nonumber \\
& \ge \frac{\gamma}{12} \sum_{n=1}^L (v(n,E))^2. \nonumber
\end{align}
Using this bound twice in \eqref{fo8}, where we take $M=kL$, we obtain
$$
P(kL, L) \ge \frac{\gamma}{12} \left( \left(\Im F(z) ||u_0(E)||_L \right)^2 +
||w(E)||_L^2 \right).
$$
Thus,
$$
h_p(z) \ge (kL)^p P(kL,L)\ge \frac{\gamma k^p}{12} \, L^p \left( \left( \Im F(z)
||u_0(E)||_L \right)^2 + ||w(E)||_L^2 \right)
$$
for any $L$ such that  $\beta \le \gamma/24$.

Next, we note that $||u_0(E)||_L^2 \ge f(L,E) = \inf_\theta ||u_\theta (E)||_L^2$ since
$u_0$ is a particular solution, corresponding to $\theta=0$. Therefore,
\begin{equation}\label{fo10}
h_p(z) \ge \frac{\gamma k^p}{12} \,  L^p (\Im F(z))^2 f(L,E).
\end{equation}
Condition (c), which, as noted above, is a consequence of the assumptions (a) and (b),
implies
\begin{equation}\label{fo11}
f(L,E) \ge D^2 L^{2 \kappa}.
\end{equation}
Up to now, we have not fixed the value of $L$. The only condition is that
$\beta \le \gamma/24$, which holds if
$$
9 \e^2 C L^{2 \alpha+1} f(L,E) \le \left( \frac{\gamma}{24} \right)^2.
$$
Since $L^{2\alpha+1} f(L,E)$ is a monotone continuous function of $L$, there exists
$L_\mathrm{max}$ (depending on $E,\e$) such that
\begin{equation}\label{fo12}
9 \e^2 C L_\mathrm{max}^{2\alpha+1} f(L_\mathrm{max}, E) = \left( \frac{\gamma}{24} \right)^2.
\end{equation}
Assume first that $p \le 2 \alpha+1$. It follows from \eqref{fo11} that
$$
D(E,\e) := \frac{f(L_\mathrm{max}, E)}{L_\mathrm{max}^{2\kappa}} \ge D^2>0.
$$
The identity \eqref{fo12} yields
$$
\e^2 L_\mathrm{max}^{2 \alpha+1+2\kappa} D(E, \e) = \frac{\gamma^2}{5184 \, C} =: \tau > 0.
$$
Thus,
$$
L_\mathrm{max}=\left( \tau \e^{-2} D^{-1} (E, \e) \right)^\frac{1}{2 \alpha+1+2 \kappa}.
$$
Inserting this expression in \eqref{fo10} with $L=L_\mathrm{max}$, we get
\begin{align*}
h_p(z) & \ge \frac{\gamma k^p}{12} \, (\Im F(z))^2 L_\mathrm{max}^{p+2 \kappa} D(E,\e) \\
& \ge \left[ \frac{\gamma k^p}{12} \, \tau^\frac{p+2 \kappa}{2 \alpha + 1 + 2 \kappa}
\right] (\Im F(z))^2 \e^\frac{-2(p+2\kappa)}{2\alpha + 1 + 2\kappa} D(E,\e)^{1-
\frac{p+2\kappa}{2\alpha+1+2 \kappa}}.
\end{align*}
Since $p \le 2 \alpha+1$, and $D(E, \e)\ge D^2>0$, we finally obtain
\begin{equation}\label{fo13}
h_p(z) \ge \text{const}(p,C,\alpha,D,\kappa,k,\gamma) \, (\Im F(z))^2
\e^{-\frac{2(p+2\kappa)}{2\alpha+1+2\kappa}}
\end{equation}
for every $z=E'+i\e$ with $E'\in A_{2\e}$, where
\begin{equation}\label{constantvalue}
\text{const}(p,C,\alpha,D,\kappa,k,\gamma) = \frac{\gamma k^p}{12} \, \left(
\frac{\gamma^2}{5184 \, C} \right)^\frac{p+2 \kappa}{2 \alpha + 1 + 2 \kappa} D^{2-
\frac{2(p+2\kappa)}{2\alpha+1+2 \kappa}}.
\end{equation}
The Parseval formula \eqref{pars} implies
$$
\langle |X|_{\delta_1}^p \rangle (T) = \frac{\e}{\pi} \int_{\R} dE' h_p(E'+i\e), \; \e =
\frac1T.
$$
Integrating only over the set $A_{2\e}$, we obtain \eqref{fa1}. The bound \eqref{fa2}
follows since $\langle |X|^p \rangle (T)$ are increasing functions of $p$.

In the case $p>2\alpha +1$, one again defines $L_\mathrm{max}$ by \eqref{fo12}. From the
(very rough) estimate $f(L_\mathrm{max},E) \le C L_\mathrm{max}^{2 \alpha+1}$, we get
$$
L_\mathrm{max} \ge \left( \frac{\gamma^2}{5184 \, C^2} \right)^\frac{1}{2}
\e^{-\frac{1}{2\alpha+1}}.
$$
The bounds \eqref{fo10} and \eqref{fo11} thus imply
$$
h_p(z) \ge \left[ \frac{\gamma k^p}{12} \left( \frac{\gamma^2}{5184 \, C^2} \right)^\frac{p
+ 2 \kappa}{2} D^2 \right] (\Im F(z))^2 \e^{-\frac{p+2\kappa}{2\alpha+1}},
$$
from which \eqref{fa3} follows.
\end{proof}

\section{Proof of Theorem~\ref{hp} For Whole-Line Operators}\label{ndlbw}

In this section we prove dynamical bounds for whole-line operators analogous to the ones
in the half-line case from Section~\ref{ndlb}.

Let $H$ be a discrete Schr\"odinger operator on $\ell^2(\Z)$ with potential $V$; compare
\eqref{oper}. For $z \in \C$ with $\Im z \not= 0$ and $n \ge 1$, we can write
$$
u(n,z)= \langle (H-z)^{-1} \delta_1, \delta_n \rangle = F(z) u_0(n,z)
+ G(z) u_{\pi/2}(n,z).
$$
(Recall that $u_\theta(z)$ denotes the solution to \eqref{evez} with $u_\theta (0,z) =
\sin \theta$, $u_\theta(1,z) = \cos \theta$.) Here, $F$ is the Borel transform of the
spectral measure $\mu$ associated with $H$ and $\delta_1$ and $G$ is a suitable
complex-valued function. Again we have, for $n \ge 1$,
\begin{equation}\label{parsanal}
a(n,T) = \langle |\langle e^{-itH} \delta_1, \delta_n \rangle |^2 \rangle (T) =
\frac{1}{\pi T} \int_\R \left| u(n,E + \tfrac{i}{T})\right|^2 dE.
\end{equation}
Write
$$
u(n,E) = F(z) u_0(n,E) + G(z) u_{\pi/2} (n,E).
$$
The analog of \eqref{eq2} is then given by
\begin{equation}\label{eq2anal}
u(n,z) = u (n,E) + (z-E) (K(E)u(z))(n), \ n \ge 1,
\end{equation}
where
$$
(K(E) \psi) (n)=\sum_{j=1}^n \left[ u_0(n,E)u_{\pi/2}(j,E) - u_{\pi/2}(n,E) u_0(j,E)
\right] \psi (j).
$$
Therefore, \eqref{eq30} holds, that is,
$$
|||K(E)|||_L^2 = \sup_\theta ||u_\theta (E)||_L^2 \inf_\theta ||u_\theta (E)||^2 .
$$
The analog of \eqref{eqa4} now reads
\begin{equation}\label{eqa4anal}
|u(n,E)|^2 = (w (n,E))^2 + (v (n,E))^2,
\end{equation}
with the two real solutions to the equation \eqref{eve},
\begin{align*}
w(n) & = \Re F(z) u_0(n,E) + \Re G(z) u_{\pi/2}(n,E), \\
v(n) & = \Im F(z) u_0(n,E) + \Im G(z) u_{\pi/2}(n,E).
\end{align*}

Let us now turn to the

\begin{proof}[Proof of Theorem~\ref{hp} for whole-line operators.]
Write $\e = 1/T$ and
$$
h_p(E'+i\e)=\sum_{n \ge 1} n^p |u(n,E'+i\e)|^2
$$
with $E' \in A_{2\e}$. Choose $E \in A$ such that $|E'-E|\le 2\e$. Now we can mimic the
proof in the half-line situation all the way up to \eqref{fo7}, which reads
\begin{align}\label{fo8anal}
P(M,L) & \ge \frac{1}{(1+\beta)^2} \sum_{n=M+1}^L |u(n,E)|^2 + \left(
(1+\beta)^{-2}-(1-\beta)^{-2} \right)||u(E)||_M^2 \\
& \ge \frac{1}{2} \sum_{n=M+1}^L |u(n,E)|^2 - 8 \beta ||u(E)||_M^2 \nonumber \\
& \ge \frac{1}{2} \left[ \sum_{n=M+1}^L (w (n,E))^2 - 8 \beta ||w(E)||_M^2 \right] +
\nonumber \\
& \qquad + \frac{1}{2} \left[ \sum_{n=M+1}^L (v (n,E))^2 - 8 \beta ||v(E)||_M^2 \right],
\nonumber
\end{align}
where we used \eqref{eqa4anal} in the last step.

Using \eqref{kboundg} twice in \eqref{fo8anal} with $M=kL$, we obtain
\begin{align*}
P(kL, L) & \ge \frac{\gamma}{12} \left( ||w(E)||_L^2 + ||v(E)||_L^2 \right) \\
& \ge \frac{\gamma}{12} \left( (v(0,E))^2 + (v(1,E))^2 \right) f(L,E) \\
& \ge \frac{\gamma}{12} \left( \Im F(E'+i\e) \right)^2 f(L,E).
\end{align*}
Thus,
$$
h_p(z) \ge (kL)^p P(kL,L)\ge \frac{\gamma k^p}{12} \, L^p \left( \Im F(E'+i\e) \right)^2
f(L,E)
$$
for any $L$ such that  $\beta \le \gamma/24$.

From this point on, we can follow the reasoning from the half-line proof and finally
obtain
$$
h_p(z) \ge \text{const}(p,C,\alpha,D,\kappa,k,\gamma) \, (\Im F(z))^2 \e
^{-\frac{2(p+2\kappa)}{2\alpha+1+2\kappa}},
$$
where $E'\in A_{2\e}$, $z=E'+i\e$ and $\text{const}(p,C,\alpha,D,\kappa,k,\gamma)$ is as
in \eqref{constantvalue}. Using \eqref{parsanal}, this allows us to conclude the proof as
before.
\end{proof}

\section{Proof of Theorem~\ref{combi}}\label{combiproof}

In this section we consider operators on the half-line and on the whole line
simultaneously and prove Theorem~\ref{combi}. Thus, let $H$ be the operator on
$\ell^2(\Z)$ or $\ell^2(Z_+)$ given by \eqref{oper}. Throughout this section, $\mu$ will
denote the spectral measure of the vector $\delta_1$ with respect to $H$, and $F$ will
denote its Borel transform,
$$
F(z) = \langle (H-z)^{-1} \delta_1, \delta_1 \rangle = \int_{\R} \frac{d \mu (x)}{x-z}.
$$

We note that all implicit constants below are positive and, if not universal, depend only
on $p$.

\begin{proof}[Proof of Theorem~\ref{combi}.]
Let $T$ be given and write $\e = 1/T$. We use \cite[Lemma~2.1]{T}. Its proof implies that
for any set A with $\mu (A) > 0$,
\begin{equation}\label{fa40}
\langle |X|_{\delta_1}^p \rangle (T) \gtrsim (\mu (A))^{1+2p} \left(J \left( A, 2 \e
\right) \right)^{-p},
\end{equation}
where
$$
J(\Delta, \nu)=\int_\Delta \int_{\R}  \frac{\nu^2}{\nu^2+(x-y)^2} \, d \mu (y) \, d \mu
(x).
$$
The fact that $2\e$ occurs in \eqref{fa40}, and not $\e$ as in \cite{T}, is due to the
fact that we use a different definition of time-averaging.

Next, one can bound $J$ from above by $I$ following the proof of \cite[Lemma~2.2]{T}. For
any $\delta>0$,
\begin{equation}\label{fa4}
I(A_{\delta}, \delta) = \int_{\R} \int_{\R} f(x,u, \delta) \, d \mu (u) \, d \mu (x) \ge
\int_A \int_{\R} f(x,u, \delta) \, d \mu (u) \, d \mu (x) ,
\end{equation}
where
$$
f(x,u,\delta) := \delta^3 \int_{A_{\delta}} \frac{dE}{((u-E)^2+\delta^2)
((x-E)^2+\delta^2)}.
$$
If $x \in A$, then $[x-\delta, x+\delta] \subset A_{\delta}$, and thus
\begin{align*}
f(x,u,\delta) & \ge \delta^3 \int_{x-\delta}^{x+\delta} \frac{dE}{((u-E)^2+\delta^2)
((x-E)^2+\delta^2)} \\
& = \int_{-1}^1 \frac{dt}{(t^2+1)((t+s)^2+1)} \\
& \gtrsim \frac{1}{s^2+1},
\end{align*}
where $s=(x-u)/\delta$. Inserting this bound into \eqref{fa4}, we get
\begin{equation}\label{fa6}
I(A_{\delta}, \delta) \gtrsim J(A, \delta).
\end{equation}
One can easily see that $I(\Delta, \nu) \gtrsim I(\Delta, 2 \nu)$. Therefore, \eqref{fa6}
implies
$$
I(A_{2\e}, \e) \gtrsim I \left( A_{2\e}, 2\e \right) \gtrsim J(A, 2\e).
$$
This, in combination with \eqref{fa40}, yields
\begin{equation}\label{fa5}
\langle |X|_{\delta_1}^p \rangle (T) \gtrsim (\mu (A))^{1+2p} \left(I (A_{2\e}, \e)
\right)^{-p}.
\end{equation}
Together with \eqref{fa30}, we obtain
$$
\langle |X|_{\delta_1}^p \rangle (T) \gtrsim \left( g_p(T) X + (\mu (A))^{1+2p} X^{-p}
\right),
$$
where $X=I(A_{2\e}, \e), \ \e=1/T$. Since  the function
$$
f(X)=aX+bX^{-p}, \; X > 0
$$
is bounded from below by
$$
\left[ p^{\frac{1}{p+1}} + p^{-\frac{p}{p+1}} \right] a^{\frac{p}{p+1}}
b^{\frac{1}{p+1}},
$$
the statement of the theorem follows.
\end{proof}

\section{Bounds for Outside Probabilities}\label{outprob}

In this section we make a connection between lower bounds for moments and lower bounds
for outside probabilities,
$$
P (|n| \ge K(T), T) = \sum_{|n| \ge K(T)} a(n,T),
$$
with an increasing function $K(T)$. We then use this to derive lower bounds for the
latter from Theorem~\ref{mome}. Thus, we can control the polynomially decaying tails of
the wave-packet (for a more detailed discussion; see \cite{GKT}).

\begin{lemma}\label{PM}
Suppose that for some $p > 0$, we have
\begin{equation}\label{p36}
\langle |X|_{\delta_1}^p \rangle (T) \ge f_p(T)
\end{equation}
with a function $f_p$ satisfying $\lim_{T \to \infty} f_p(T) =
\infty$. Then, for any $\delta>0$, we have
$$
P \left( |n| \ge \left( \frac{f_p(T)}{2} \right)^\frac{1}{p} , T \right) \gtrsim
T^{-p(1+\delta)} f_p(T).
$$
\end{lemma}

\begin{proof}
Write $K_p(T) = \left( f_p(T)/2 \right)^{1/p}$. For given
$\delta>0$, consider the following sets,
\begin{align*}
S_1 & = \{ n \in \Z : |n| \le K_p(T) \}, \\
S_2 & = \{ n \in \Z : K_p(T) < |n| \le T^{1+\delta} \}, \\
S_3 & = \{ n \in \Z : |n| > T^{1+\delta} \},
\end{align*}
and denote by $A_1,A_2,A_3$ the corresponding partial sums in the
definition of $\langle |X|_{\delta_1}^p \rangle (T)$, that is,
$$
A_j = \sum_{n \in S_j} |n|^p a(n,T), \;\; 1 \le j \le 3.
$$

Since for every $T$, $\sum_n a(n,T)=1$, it follows that
\begin{equation}\label{p38}
A_1 \le K_p^p(T) = \frac{f_p(T)}{2}.
\end{equation}
For $A_2$, we have the obvious bound
\begin{equation}\label{p39}
A_2 \le T^{p(1+\delta)} P(|n| \ge K_p (T), T).
\end{equation}
For every $s > 0$, we can estimate $A_3$ as follows:
\begin{align*}
A_3 & = \sum_{|n|>T^{1+\delta}} |n|^p a(n,T) \\
& \le T^{-s(1+\delta)} \sum_{|n| >T^{1+\delta}} |n|^{p+s} a(n,T) \\
& \le T^{-s(1+\delta)} \langle |X|_{\delta_1}^{p+s} \rangle (T).
\end{align*}
As there is a ballistic upper bound for the moments, that is, $\langle |X|_{\delta_1}^r
\rangle (T) \le C(r)T^r$, we obtain, taking $s = p/\delta$,
\begin{equation}\label{p40}
A_3 \le C,
\end{equation}
with a $T$-independent constant $C$. The bounds
\eqref{p36}--\eqref{p40} yield
\begin{align*}
T^{p(1+\delta)} P(|n| \ge K_p(T),T) & \ge A_2 \\
& \ge \langle |X|_{\delta_1}^p \rangle (T) - A_1 - A_3 \\
& \ge \frac{f_p(T)}{2} - C.
\end{align*}
Since $\lim_{T \to \infty} f_p (T) = \infty$, the lemma follows.
\end{proof}

\noindent\textit{Remark.} The result holds, of course, for any
well-localized initial state $\psi, \ ||\psi||=1$, not necessarily
$\delta_1$.

\medskip

Under the conditions of Theorem~\ref{mome} this lemma gives lower
bounds of the form $ P(|n| \ge T^\gamma, T) \ge T^{-g(\gamma)}$,
where
$$
\gamma_1 = \frac{2\kappa}{\alpha+\kappa+1/2} < \gamma <
\frac{1}{\alpha+1} = \gamma_2,
$$
and $g(\gamma)$ is an explicit positive growing function. This can
be achieved by taking appropriate values of the parameter $p$. In
particular, if one takes $p$ small, then $\gamma$ is close to
$\gamma_1$ and $g(\gamma)$  is close to $0$. If $p$ is close to
$2\alpha+1$, then $\gamma$ is close to $\gamma_2$ and $g(\gamma)$ is
close to $\alpha(2 \alpha+1)/(\alpha+1)$.

\section{Application to Quasicrystal Models}\label{fibsec}

In this section we consider operators with Sturmian potentials (i.e., given by
\eqref{sturmpot}) and prove Theorem~\ref{fibthm}.

Recall that the continued fraction expansion of $\omega$ is given by
$$
\omega = \cfrac{1}{a_1+ \cfrac{1}{a_2+ \cfrac{1}{a_3 + \cdots}}},
$$
with uniquely determined $a_n \in \Z_+$. The associated rational approximants $p_n/q_n$
are defined by
\begin{alignat*}{3}
p_0 &= 0, &\quad    p_1 &= 1,   &\quad  p_n &= a_n p_{n-1} + p_{n-2},\\
q_0 &= 1, &     q_1 &= a_1, &       q_n &= a_n q_{n-1} + q_{n-2}.
\end{alignat*}
The number $\omega$ is said to have bounded partial quotients if the sequence $\{a_n\}$
is bounded. More generally, it is said to have bounded density if $d(\omega)$, as defined
in \eqref{domega}, is finite. Both sets of numbers are uncountable and have Lebesgue
measure zero.

Our goal is to establish solution estimates for all energies in the spectrum in order to
apply our general dynamical result. The upper bound is known under certain assumptions on
$\omega$. Namely, the following proposition is a consequence of \cite[Corollary~10]{irt}:

\begin{prop}\label{fib1}
Suppose $\omega$ is a bounded density number. For every $\lambda$, there is a constant
$C$ such that for every $E \in \sigma(H)$, every normalized solution $u$ of \eqref{eve},
and every $L \ge 1$, $\|u(E)\|_L^2 \le C L^{2\alpha+1}$, with
\begin{equation}\label{alphaval}
\alpha = D \, d(\omega) \, \log C_\lambda,
\end{equation}
where $D$ is some universal constant, $C_\lambda = 2 + \sqrt{8 + \lambda^2}$, and
$d(\omega)$ is as in \eqref{domega}.
\end{prop}

It remains to prove suitable lower bounds for solutions. Define the words $s_n$ over the
alphabet $\mathcal{A}=\{0,\lambda\}$ by
\begin{equation}\label{recursive}
s_{-1}^{} = \lambda, \quad s_0^{} = 0, \quad s_1^{} = s_0^{a_1 - 1} s_{-1}^{}, \quad
s_n^{} = s_{n-1}^{a_n} s_{n-2}^{}, \; n \ge 2.
\end{equation}
In particular, the word $s_n$ has length $q_n$ for each $n \ge 0$. By definition,
$s_{n-1}$ is a prefix of $s_n$ for each $n \ge 2$. Thus, the words have a one-sided
infinite limit which, in fact, coincides with the restriction of $V$ to the right
half-line (see \cite{dkl,dl1}).

For later use, we recall the following elementary formula \cite[Proposition~2.3]{dl1}
which implies that the word $s_n s_{n+1}$ has $s_{n+1}$ as a prefix:
\begin{equation}\label{wunder}
s_n^{} s_{n+1}^{}= s_{n+1}^{} s_{n-1}^{a_n - 1} s_{n-2}^{} s_{n-1}^{} \, \text{ for every
} n \ge 2.
\end{equation}

Another important ingredient is that for each energy $E$ in the spectrum, the trace
$x_n(E)$ of the transfer matrix $T(q_n,0;E)$ from $0$ to $q_n$ obeys
\begin{equation}\label{tm}
\min\{ |x_n(E)|, |x_{n+1}(E)| \} \le 2.
\end{equation}
This was shown by Bellissard et al.\ in \cite{bist} (see also S\"ut\H{o} \cite{su} for
this result in the Fibonacci case).

Such trace bounds are useful as shown by the following lemma, which has been used a
number of times \cite{d,dkl,JL2,KKL}. Given a solution $u$ to \eqref{eve}, we write $U(n)
= (u(n+1,E),u(n,E))^T$ for the associated $2$-vector. Thus, $U(n) = T(n,0;E) U(0)$ for
every $n$. We define $\|U(n)\|^2 = |u(n)|^2 + |u(n+1)|^2$ and, as before,
$$
\|U\|_L^2 = \sum_{n=1}^{[L]} \|U(n)\|^2 + (L-[L]) \|U([L]+1)\|^2.
$$

\begin{lemma}\label{squares}
Suppose $p,q \in \Z_+$ are such that $p \ge q$ and $V(m + p) = V(m)$ for $1 \le m \le
p+q$. Then, we have
$$
\|U\|^2_{2p+q} \ge \left( 1 + \left( \frac{1}{\max \{ 2, 2|\tr T(p,0;E)|\} }\right)^2
\right) \|U\|^2_q
$$
for every solution $u$ to \eqref{eve}.

In particular, if $|\tr T(p,0;E)| \le 2$, then
$$
\|U\|^2_{2p+q} \ge \frac{17}{16} \, \|U\|^2_q
$$
for every solution $u$ to \eqref{eve}.
\end{lemma}

\begin{proof}
As mentioned above, this lemma is known. However, for the convenience of the reader, we
supply the short proof.

By the assumption, the cyclicity of the trace, and the Cayley-Hamilton theorem,
$$
U(2p+m) - \tr T(p,0;E) \, U(p+m) + U(m) = 0
$$
and hence
$$
\max \left\{ \|U(p+m)\| , \|U(2p+m)\| \right\} \ge \frac{1}{\max \{ 2, 2|\tr T(p,0;E)|\}}
\|U(m)\|
$$
for all $1 \leq m \leq q$. We can therefore proceed as follows,
\begin{align*}
\|U\|_{2p+q}^2 &= \sum_{m=1}^{2p+q} \|U(m)\|^2\\
& =   \sum_{m=1}^{q} \|U(m)\|^2\; + \sum_{m=q+1}^{2p+q} \|U(m)\|^2\\
& \ge \sum_{m=1}^{q} \|U(m)\|^2 + \left( \tfrac{1}{\max \{ 2, 2|\tr T(p,0;E)|\}} \right)^2 \sum_{m=1}^{q} \|U(m)\|^2\\
& =   \left( 1 + \left( \tfrac{1}{\max \{ 2, 2|\tr T(p,0;E)|\}} \right)^2 \right)
\|U\|_q^2.
\end{align*}
This proves the assertion.
\end{proof}

With these tools at our disposal, we can prove the following scaling result for solutions
along a sequence of the form $\{q_{5n + n_0}\}$.

\begin{lemma}\label{gordon}
For every $\lambda > 0$, $\omega \in (0,1)$ irrational, $E \in \sigma(H)$, and every
normalized solution $u$ of \eqref{eve}, we have
$$
\|U\|^2_{q_{n+5}} \ge \frac{17}{16} \, \|U\|^2_{q_n}
$$
for all $n \ge 0$.
\end{lemma}

\begin{proof}
By \eqref{wunder}--\eqref{tm} and Lemma~\ref{squares}, we only need to produce two
\textit{consecutive} squares followed be a suitable prefix within five levels of the
$s_n$-hierarchy.

Consider first the case $a_{n+5} \ge 2$:
\begin{align*}
s_{n+5} & = s_{n+4}^2 s_{n+3} \ldots\\
& = s_{n+3}^2 s_n \ldots,
\end{align*}
where one possibly has to use \eqref{wunder}. This yields two consecutive squares and we
can now apply Lemma~\ref{squares} with trace bound $2$.

Now, consider the case $a_{n+5} = 1$:
\begin{align*}
s_{n+5} & = s_{n+4} s_{n+3}\\
& = s_{n+3}^{a_{n+4}} s_{n+2} s_{n+3} = s_{n+3}^{a_{n+4} + 1} s_{n+1}^{a_{n+2} - 1} s_n
s_{n+1} \\
& = ( s_{n+2}^{a_{n+3}} s_{n+1} )^{a_{n+4}} s_{n+2} s_{n+3}.
\end{align*}
Now, if $a_{n+2} \ge 2$, we are done. Otherwise, we apply \eqref{wunder} twice and find a
suitable square.
\end{proof}

This shows that along the $q_n$ scales, we find suitable exponential growth of solutions.
If the $q_n$'s have reasonable growth, this translates into nice bounds for all values of
$L$.

\begin{prop}\label{fib2}
Suppose $\lambda > 0$ and $\alpha \in (0,1)$ is irrational with $a_n \le C$, $\omega =
0$. Consider the operator \eqref{oper} with potential given by \eqref{sturmpot}. Then,
for every $E \in \sigma(H)$ and every normalized real solution $u$ of
\eqref{eve}, the following hold true:\\
{\rm (a)} With $k = (C + 1)^{-6} \in (0,1)$ and $\gamma = 1/16$, we have for $L$ large
enough,
$$
\|U\|^2_L \ge (1 + \gamma) \|U\|^2_{kL}.
$$
{\rm (b)} With
\begin{equation}\label{kappaval}
\kappa = \frac{\log (\sqrt{17} / 4)}{(C+1)^5}
\end{equation}
and some {\rm (}$E$ and $u$-independent{\rm )} constant $D$, we have for $L \ge 1$,
$$
\|U\|_L \ge DL^\kappa.
$$
\end{prop}

\noindent\textit{Remark.} ``Large enough'' $L$ means, for example, $L \ge q_5$. If
$\tilde{C} = \limsup a_n < C$, one can choose in (b) a larger $\kappa$ (namely, with $C$
in \eqref{kappaval} replaced by $\tilde{C}$) if one requires the estimate only for $L \ge
L_0$; $L_0$ depends on how close to the maximum possible value one wants to choose
$\kappa$.

\begin{proof}
(a) Given $L$, define $n$ by $q_n \le L < q_{n+1}$. Then, by Lemma~\ref{gordon},
\begin{align*}
\|U\|^2_L & \ge \|U\|^2_{q_n} \ge \frac{17}{16} \|U\|^2_{q_{n-5}} = (1 + \gamma)
\|U\|^2_{q_{n-5}}\\ & \ge (1 + \gamma) \|U\|^2_{k q_{n+1}} \ge (1 + \gamma) \|U\|^2_{k
L}.
\end{align*}
(b) We know already that $\|U\|_{q_{5n}} \ge \mathrm{const} \cdot
(\frac{\sqrt{17}}{4})^n$. Define $n$ by $q_{5n} \le L < q_{5(n+1)}$. Then
\begin{align*}
\|U\|_L & \ge \|U\|_{q_{5n}} \ge \mathrm{const} \cdot \left(\frac{\sqrt{17}}{4}\right)^n
= \frac{4 \cdot \mathrm{const}}{\sqrt{17}} \left(\frac{\sqrt{17}}{4}\right)^{n+1} \\
& \ge \frac{4 \cdot \mathrm{const}}{\sqrt{17}} \left(q_{5(n+1)}\right)^\kappa \ge \frac{4
\cdot \mathrm{const}}{\sqrt{17}} L^\kappa,
\end{align*}
as claimed.
\end{proof}

\begin{proof}[Proof of Theorem~\ref{fibthm}.]
With $A = \sigma(H)$, we clearly have $\mu(A) > 0$, where, as before, $\mu$ denotes the
spectral measure of $\delta_1$ with respect to $H$. Thus, combining
Propositions~\ref{fib1} and \ref{fib2} with Theorem~\ref{mome}, we obtain the claimed
lower bound \eqref{fibbetabound} for $\beta_{\delta_1}^-(p)$. Note that $\|u\|_L$ and
$\|U\|_L$ are, of course, comparable.
\end{proof}

\noindent\textit{Remark.} We end this section with a brief discussion of the more general
potentials
\begin{equation}\label{sturmpotgen}
V(n) = \lambda \chi_{[1-\omega,1)}(n \omega + \theta \mod 1), \; n \in \Z,
\end{equation}
where $\lambda > 0$, $\omega$ is an irrational number with bounded partial quotients, and
$\theta \in [0,1)$. Essentially, these are the elements of the hulls generated by
potentials of the form \eqref{sturmpot} and are the natural objects when studying these
operators within the framework of ergodic families of operators; compare \cite{cl,pf}.

For the potentials in \eqref{sturmpotgen}, we can perform an analysis almost completely
parallel to the one above. The only difference is that it is not possible to obtain the
exact analog of Lemma~\ref{gordon}. Rather, the universal constant $17/16$ has to be
replaced by a $\lambda$-dependent constant that goes to $1$ as $\lambda \to \infty$.

The net result is that we are able to prove
$$
\beta^-_{\delta_1}(p) \ge \begin{cases} \frac{p (p+2\kappa)}{(p+1)(\alpha + \kappa + 1/2)}
& p \le 2 \alpha+1, \\
\frac{p}{\alpha+1} & p>2 \alpha +1 \end{cases}
$$
with the same $\lambda$-dependent constants $\alpha$ and $\tau$ as in \eqref{sturmdkl}.
In particular, our approach gives better dynamical results for potentials of the form
\eqref{sturmpotgen} than previous ones. However, we are limited here to the class of
$\omega$'s with bounded partial quotients, whereas \cite{dkl} could work with the
slightly larger class of bounded density numbers.

\begin{appendix}

\section{The Jitomirskaya-Last Method Revisited}\label{jlsec}

The Jitomirskaya-Last inequality, \eqref{la1} below, provides a link between two limiting
procedures by a clever association of a length scale to a given small $\e$. This
inequality immediately implies all results of Gilbert-Pearson theory \cite{gp}. Moreover,
it allows for a strengthening of this theory that is needed to study
Hausdorff-dimensional properties of spectral measures (see, e.g.,
\cite{d,dk,dkl,dl,JL,JL2,KKL,KLS,r,z} for applications in this context). A variant of the
Jitomirskaya-Last inequality was proven and further developed in \cite{KKL}; see
\eqref{la2} and \eqref{la21} below. These refinements were crucial in the approach to new
dynamical bounds in \cite{KKL}. In this appendix we recount these central results of
Jitomirskaya-Last theory and provide simplified proofs for some of them.

Let us denote
\begin{align*}
a(L) & = ||u_{\pi/2}(E)||_L^2, \\  b(L) & = ||u_0(E)||_L^2, \\
d(L) & = \langle u_0(E), u_{\pi/2}(E) \rangle _L, \\ w^2 (L) & = |||K(E)|||_L^2.
\end{align*}
From \eqref{eq30}, we see that $(w (L))^2 = ab-d^2$. Of course, all these numbers also
depend on $E$. We will leave this implicit throughout this section but remark that these
quantities are associated with the energy $E$ even in situations involving some
additional energy $E'$.

For non-negative functions $f$ and $g$, we write $f \sim g$ if we have both $f \gtrsim g$ and
$g \gtrsim f$, that is,
$$
f \sim g \; : \Leftrightarrow \; \exists \, C_1,C_2 > 0 \text{ such that } C_1 f(x) \le
g(x) \le C_2 f(x) \text{ for all } x.
$$

It was shown in \cite{JL} that
\begin{equation}\label{la1}
|F(E+i\e)|^2 \sim \frac{ a(L_1(\e)) }{ b(L_1(\e)},
\end{equation}
where $L_1(\e)$ is defined by the equality
$$
a(L_1(\e))b(L_1(\e))=\frac{1}{4\e^2}.
$$
Later on \cite[Theorem~2.3]{KKL}, it was shown that
\begin{equation}\label{la2}
|F(E+i\e)|^2 \sim \frac{ a(L_2(\e)) }{ b(L_2(\e)},
\end{equation}
where $L_2(\e)$ is defined by
$$
w(L_2(\e))=\frac{1}{\e}.
$$
Although not stated explicitly, it follows also from \cite[Theorem~2.3]{KKL} that
\begin{equation}\label{la21}
\Im F(E+i\e) \sim \frac{w(L_2(\e)) }{ b(L_2(\e))}.
\end{equation}
The constants $C_{1,2}$ in \eqref{la1}--\eqref{la21} are universal. The lower bound for
$|F(z)|$ can be used to prove the singularity of the spectral measure and the upper bound
to prove its continuity. In fact, when considering the continuity of the measure, only
the upper bound on $\Im F(E+i\e)$ matters, since one can use the inequality
$$
\mu ([E-\e, E+\e]) \le 2 \e \, \Im F(E+i\e).
$$
\par
Another result proved in \cite{KKL} that develops ideas of \cite{JL} concerns a lower
bound for the inside probabilities $\sum_{n<L(T)} a(n,T)$ with some growing $L(T)$.

We shall present below some version of the Jitomirskaya-Last method which gives, in
particular, a simplified proof of \eqref{la21}. We begin with the following technical
result.

\begin{lemma}\label{main}
Suppose that $E,E' \in \R$ and $\e > 0$. Write $z=E'+i\e$. Then, for $L>0$,
\begin{align}
\label{eq4.1} \Im F(z) & \ge \e ||u(z)||_L^2 \\
\label{eq4.2} & \ge \frac{\e}{(1+ |z-E| w(L) )^2} \left((\Im F(z))^2 b(L)+ \frac{(w(L))^2}{b(L)}\right) \\
\label{eq4.3} & \ge \frac{2\e w(L)}{(1+ |z'-E| w(L) )^2} \, \Im F(z).
\end{align}

\end{lemma}

\begin{proof}
The inequality \eqref{eq4.1} is obvious (in fact, equality holds for $L = \infty$). From
\eqref{eqa3} and \eqref{eqa4}, we obtain
\begin{align*}
(1+|z-E| w(L))^2 ||u(z)||_L^2 & \ge a(L)+((\Im F(z))^2 + (\Re F(z))^2) b(L) -2 \Re F(z) \, d(L)\\
 & \ge (\Im F(z))^2 b(L) + (a(L)b(L)-d^2(L))/b(L) \\
 & = (\Im F(z))^2 b(L) + w^2(L)/b(L),
\end{align*}
which implies \eqref{eq4.2}. (The second step above uses $x^2b -2xd \ge -d^2/b$ for any
$x \in \R$.) Finally, \eqref{eq4.3} follows from the elementary bound $b s^2+w^2/b \ge 2
ws, \ b>0$.
\end{proof}

As a first consequence of this lemma, one can obtain the equivalence \eqref{la21} for
$\Im F(E+i\e)$.

\begin{prop}\label{U}
The following inequalities hold:
\begin{align}
\label{boU} \Im F(E+i\e) & \le \inf_L  \frac{(1+ \e w(L))^2 }{ \e b(L)} \le \frac{4
w(L_2(\e)) }{ b(L_2(\e))}, \\
\label{boL} \Im F(E+i\e) & \ge \sup_L \frac{\e w^2(L) }{ b(L)(1+\e w(L))^2} \ge
\frac{w(L_2(\e)) }{ 4 b(L_2(\e))}.
\end{align}
\end{prop}

\begin{proof}
The bounds \eqref{eq4.1}--\eqref{eq4.2} with $E'=E$ yield
$$
\Im F (E+i\e) \ge \e \frac{ (\Im F(E+i\e))^2 b(L)}{(1+\e w(L))^2}.
$$
Thus,
$$
\Im F(E+i\e) \le \frac{(1+\e w(L))^2 }{ \e b(L)}
$$
for any $L$ and in particular for $L=L_2(\e)$. The bound \eqref{boU} follows. On the
other hand, again from \eqref{eq4.1}--\eqref{eq4.2},
$$
\Im F(E+i\e) \ge \frac{\e w^2(L) }{ b(L) (1+\e w(L))^2}
$$
for any $L$, which yields \eqref{boL}.
\end{proof}

Another consequence is the following result, which is Proposition~2.4 of \cite{KKL}.
Together with the Parseval identity, it yields a lower bound for the time-averaged inside
probabilities; see Theorem~1.1 and its proof in \cite{KKL}. The proof we give is simpler
than the one in \cite{KKL} and provides a better constant on the right-hand side (where
we have $\sqrt{2}/4$, they have $(3-2\sqrt{2})/36$). Moreover, it is clear why should one
take $L$ large enough but not too large.

\begin{prop}\label{lowerin}
Suppose $E,E' \in \R$, $\e > 0$, and $|E - E'| < \e$.  Write $z=E'+i\e$ and define
$L_3(\e)$ by
$$
\sqrt{2} \e w(L_3(\e))=1.
$$
Then
$$
\e ||u(z)||^2_{L_3(\e)} \ge \frac{\sqrt{2}}{4} \, \Im F(z).
$$
\end{prop}

\noindent\textit{Remark.} Note that both $u$ and $F$ are associated with $z = E' + i\e$,
but the length scale $L_3(\e)$ is defined using quantities associated with energy $E$.
This fact is important to the applications of this result; compare the remark after
\cite[Proposition~2.4]{KKL} and the proof of \cite[Theorem~1.1]{KKL}.

\begin{proof}
Clearly, $|z-E| \le \sqrt{2}\e$. It follows from \eqref{eq4.2}--\eqref{eq4.3} that
$$
\e ||u(z)||_L^2 \ge \frac{2\e w(L)}{(1+  \sqrt{2} \e w(L))^2} \, \Im F(z)
$$
for any $L$. Since the function $f(y)=\frac{2y}{(1+\sqrt{2}y)^2}, \ y>0$, has its maximum
at $y_0 = \frac{1}{\sqrt{2}}$, the result follows.
\end{proof}

\end{appendix}


\begin{thebibliography}{BIST}

\bibitem[BIST]{bist} J.\ Bellissard, B.\ Iochum, E.\ Scoppola, and D.\ Testard, Spectral properties
of one-dimensional quasicrystals, \textit{Commun.\ Math.\ Phys.}\ {\bf 125} (1989),
527--543

\bibitem[B]{b} Yu.\ M.\ Berezanski\u\i, \textit{Expansions in Eigenfunctions of Selfadjoint Operators},
Transl.\ Math.\ Monographs {\bf 17}, American Mathematical Society, Providence (1968)

\bibitem[CL]{cl} R.\ Carmona and J.\ Lacroix, \textit{Spectral Theory of Random Schr\"odinger
Operators}, Birkh\"auser, Boston (1990)

\bibitem[CM]{cm} J.-M.\ Combes and G.\ Mantica, Fractal dimensions and quantum evolution associated
with sparse potential Jacobi matrices, in \textit{Long Time Behaviour of Classical and
Quantum Systems} (\textit{Bologna, 1999}), pp.~107--123, Ser.\ Concr.\ Appl.\ Math. {\bf
1}, World Sci.\ Publishing, River Edge (2001)

\bibitem[D]{d} D.\ Damanik, $\alpha$-continuity properties of one-dimensional quasicrystals,
\textit{Commun.\ Math.\ Phys.} {\bf 192} (1998), 169--182

\bibitem[DK]{dk} D.\ Damanik and R.\ Killip, Half-line Schr\"odinger operators with no bound states,
Preprint (mp-arc/03-77), to appear in \textit{Acta Math.}

\bibitem[DKL]{dkl} D.\ Damanik, R.\ Killip, and D.\ Lenz, Uniform spectral properties of one-dimensional
quasicrystals. III. $\alpha$-continuity, \textit{Commun.\ Math.\ Phys.}\ {\bf 212}
(2000), 191--204

\bibitem[DLa]{dl} D.\ Damanik and M.\ Landrigan, Log-dimensional spectral properties of one-dimensional quasicrystals,
\textit{Proc.\ Amer.\ Math.\ Soc.} {\bf 131} (2003), 2209--2216

\bibitem[DLe]{dl1} D.\ Damanik and D.\ Lenz, Uniform spectral properties of
one-dimensional quasicrystals, I. Absence of eigenvalues, \textit{Commun.\ Math.\ Phys.}\
{\bf 207} (1999), 687--696

\bibitem[DST]{DST} D.\ Damanik, A.\ S\"ut\H{o}, and S.\ Tcheremchantsev, Power-Law bounds on transfer
matrices and quantum dynamics in one dimension II., Preprint (mp-arc/03-52), to appear in
\textit{J.\ Funct.\ Anal.}

\bibitem[DT]{DT} D.\ Damanik and S.\ Tcheremchantsev, Power-Law bounds on transfer matrices and
quantum dynamics in one dimension, \textit{Commun.\ Math.\ Phys.} {\bf 236} (2003),
513--534

\bibitem[GKT]{GKT} F.\ Germinet, A.\ Kiselev, and S.\ Tcheremchantsev, Transfer matrices and transport for
Schr\"odinger operators, Preprint (mp-arc/03-240), to appear in \textit{Ann.\ Inst.\
Fourier}

\bibitem[GP]{gp} D.\ J.\ Gilbert and D.\ B.\ Pearson, On subordinacy and analysis of the
spectrum of one-dimensional Schr\"odinger operators, \textit{J.\ Math.\ Anal.\ Appl.}
{\bf 128} (1987), 30--56

\bibitem[IRT]{irt} B.\ Iochum, L.\ Raymond, and D.\ Testard, Resistance of one-dimensional
quasicrystals, \textit{Physica A} {\bf 187} (1992), 353--368

\bibitem[JL1]{JL} S.\ Jitomirskaya and Y.\ Last, Power-law subordinacy and singular spectra.
I. Half-line operators, \textit{Acta Math.} {\bf 183} (1999), 171--189

\bibitem[JL2]{JL2} S.\ Jitomirskaya and Y.\ Last, Power-law subordinacy and singular spectra.
II. Line operators, \textit{Commun.\ Math.\ Phys.} {\bf 211} (2000), 643--658

\bibitem[JSS]{JSS} S.\ Jitomirskaya, H.\ Schulz-Baldes, and G.\ Stolz, Delocalization in random
polymer models, \textit{Commun.\ Math.\ Phys.} {\bf 233} (2003), 27--48

\bibitem[KKL]{KKL} R.\ Killip, A.\ Kiselev, and Y.\ Last, Dynamical upper bounds on wavepacket
spreading, \textit{Amer.\ J.\ Math.} {\bf 125} (2003), 1165--1198

\bibitem[KLS]{KLS} A.\ Kiselev, Y.\ Last, and B.\ Simon, Stability of singular spectral types
under decaying perturbations, \textit{J.\ Funct.\ Anal.} {\bf 198} (2003), 1--27

\bibitem[PF]{pf} L.\ Pastur and A.\ Figotin, \textit{Spectra of Random and Almost-Periodic Operators},
Grundlehren der Mathematischen Wissenschaften {\bf 297}, Springer-Verlag, Berlin (1992)

\bibitem[RS]{rs4} M.\ Reed and B.\ Simon, \textit{Methods of Modern Mathematical Physics.
IV.~Analysis of Operators}, Academic Press, New York-London (1978)

\bibitem[R]{r} C.\ Remling, The absolutely continuous spectrum of one-dimensional Schr\"odinger
operators with decaying potentials, \textit{Commun.\ Math.\ Phys.} {\bf 193} (1998),
151--170

\bibitem[Si1]{Si1} B.\ Simon, Schr\"odinger semigroups, \textit{Bull.\ Amer.\ Math.\ Soc.} {\bf 7} (1982),
447--526

\bibitem[Si2]{Si} B.\ Simon, Bounded eigenfunctions and absolutely continuous spectra for
one-dimensional Schr\"odinger operators, \textit{Proc. Amer. Math. Soc.} {\bf 124}
(1996), 3361--3369

\bibitem[S\"u]{su} A.\ S\"ut\H{o}, The spectrum of a quasiperiodic Schr\"odinger operator, \textit{Commun.\
Math.\ Phys.}\ {\bf 111} (1987), 409--415

\bibitem[T1]{T2} S.\ Tcheremchantsev, Mixed lower bounds for quantum transport, \textit{J.\ Funct.\
Anal.} {\bf 197} (2003), 247--282

\bibitem[T2]{T} S.\ Tcheremchantsev, Dynamical analysis of Schr\"odinger operators with growing sparse
potentials, Preprint (mp-arc/03-472), to appear in \textit{Commun.\ Math.\ Phys.}

\bibitem[Z]{z} A.\ Zlato\v s, Sparse potentials with fractional Hausdorff dimension, \textit{J.\
Funct.\ Anal.} {\bf 207} (2004), 216--252

\end{thebibliography}
\end{document}